\documentclass[10pt]{emulateapj}
\usepackage{float}

\def\be{\begin{equation}}
\def\ee{\end{equation}}
\def\bea{\begin{eqnarray}}
\def\eea{\end{eqnarray}}
\def\f{\frac}

\shorttitle{GRB Spectrum with Decaying Magnetic Field}
\shortauthors{Zhao et al.}

\begin{document}
\title{Gamma-Ray Burst Spectrum with Decaying Magnetic Field}
\author{Xiaohong Zhao\altaffilmark{1,2,4}, Zhuo Li\altaffilmark{3,2}, Xuewen Liu\altaffilmark{4,5},
Binbin Zhang\altaffilmark{4}, Jinming Bai\altaffilmark{1,2} and Peter
M\'{e}sz\'{a}ros\altaffilmark{4}} \altaffiltext{1}{National
Astronomical Observatories/Yunnan Observatory, Chinese Academy of
Sciences, P.O. Box 110, 650011 Kunming, China; zhaoxh@ynao.ac.cn}
\altaffiltext{2}{Key Laboratory for the Structure and Evolution of Celestial Bodies,
Chinese Academy of Sciences, P.O. Box 110, 650011 Kunming, China }\altaffiltext{3}{Department of Astronomy and Kavli Institute for
Astronomy and Astrophysics, Peking University, Beijing 100871,
China; zhuo.li@pku.edu.cn} \altaffiltext{4} {Department of Astronomy
\& Astrophysics and Department of Physics, The Pennsylvania State University, University
Park, PA 16802, USA} \altaffiltext{5} {Department of Physics,
Sichuan University, Chengdu 610065, China}

\begin{abstract}
In the internal shock model for gamma-ray bursts (GRBs), the
synchrotron spectrum from the fast cooling electrons in a
homogeneous downstream magnetic field (MF) is too soft to produce
the low-energy slope of GRB spectra. However the magnetic field may
decay downstream with distance from the shock front. Here we show
that the synchrotron spectrum becomes harder if electrons undergo
synchrotron and inverse-Compton cooling in a decaying MF. To
reconcile this with the typical GRB spectrum with low energy slope
$\nu F_\nu\propto\nu$, it is required that the postshock MF decay
time is comparable to the cooling time of the bulk electrons
(corresponding to a MF decaying length typically of $\sim10^5$ skin
depths); that the inverse-Compton cooling should dominate
synchrotron cooling after the MF decay time; and/or that the MF
decays with comoving time roughly as $B\propto t^{-1.5}$. An
internal shock synchrotron model with a decaying MF can account for
the majority of GRBs with low energy slopes not harder than
$\nu^{4/3}$.
\end{abstract}


\section{Introduction}
The radiation mechanism of gamma-ray bursts (GRBs) is still one of
the key problems in GRB physics. The GRB spectra usually can be well
fit by the Band function (Band et al. 1993), where two power law
sections are smoothly jointed. The low and high energy photon
indices are typically $\alpha\sim-1$ and $\beta\sim-2.2$,
respectively, and the $\nu F_\nu$ spectral peak energy is $E_p\sim250$~keV.
(Preece et al. 2000). Given the
non-thermal spectra and the high luminosity of GRBs, it is widely
believed that the main radiation mechanisms at work are synchrotron
and/or inverse-Compton (IC) radiations (M\'{e}sz\'{a}ros et al.
1994, Tavani 1996, Cohen et al. 1997). Since the spectral bump in
sub-MeV range dominates the energy flux, synchrotron is more favored
over IC to account for the sub-MeV emission
\citep{derishev01,piran09}.

However, more careful studies have raised questions about the simple
synchrotron model. In the widely used internal shock model (Rees \&
M\'{e}sz\'{a}ros 1994), the energy dissipation in GRBs is caused by
collisions between different parts of the unsteady outflow. These
collisions produce shocks which accelerate electrons and generate
magnetic field, and the GRB prompt emission is produced by the
synchrotron radiation from the accelerated electrons. The high
energy spectral slope $\beta\sim-2.2$ is consistent with the
synchrotron spectrum from fast cooling electrons injected as a
power law energy distribution with the particle index $p\sim2.3$, a
typical value in Fermi shock acceleration. However, in order to
produce synchrotron photon of $E_p$ at $\sim$sub-MeV range, the
magnetic field strength should be strong, close to equipartition
value. The strong fast cooling of electrons in strong magnetic field
produces a low energy spectral slope with $\alpha=-3/2$, extending
from the injection energy $E_p$ down to very low energy, much softer
than observed ones. This raises the fast cooling problem of
synchrotron radiation in GRBs (Ghisellini et al. 2000, Preece et al. 2000).

In recent years an alternative model based on photospheric emission
has been widely discussed for explaining the prompt emission (M\'{e}sz\'{a}ros \&
Rees, 2000; Rees \& M\'{e}sz\'{a}ros 2005; Ryde 2005; Pe'er et al.
2006; Beloborodov 2010). In general a hard spectrum, as hard as $\nu
F_\nu\propto\nu^{3}$, may be produced. A number of GRBs are found to
be consistent with photospheric emission (e.g., Ryde et al. 2010; Pe'er, et
al. 2012). However, the spectrum predicted can be too hard at
low energies and too soft at high energies, compared with typical
spectral slopes of $\alpha\sim-1$ and $\beta\sim-2.2$. So a
non-thermal emission component may still play a crucial role in the prompt
emission.

An underlying assumption in the traditional synchrotron internal
shock model is that the downstream magnetic field (MF) is
homogeneous. However, as discussed by Gruzinov \& Waxman (1999) and
Gruzinov (2001) in the case of afterglow shocks, if the MF is
generated by the Weibel instability, the MF would maintain an
equipartition value only within a skin depth of the plasma,
$c/\omega_p$, where $\omega_p$ is the proton  plasma frequency. The detailed
processes of particle acceleration and MF formation in collisionless
shock are still unclear, but numerical simulations are making
progress. Recent simulations of shocks indicate that the Weibel
instability induced filaments merge and cause the MF to gradually
decay (e.g., Chang et al. 2008; \cite{Silva03}; \cite{Medv05}). The
particle-in-cell (PIC) simulation of Chang et al. (2008) indicates
that the MF decays as a power law of time. The longer PIC simulation
by Keshet et al. (2009) up to $\sim10^4w_p^{-1}$ seems to suggest an
exponential decay to $\epsilon_B\sim10^{-2}$ at a few hundreds skin
depths from the shock front. However, all the simulations currently
only probe a region much smaller than the shocked region. Thus the MF
evolution on longer time or space scales is still unknown.

Given the uncertainty in MF evolution behind shocks, Rossi \& Rees
(2003) and Lemoine (2013) have investigated the effect of MF decay
on afterglow emission. Lemoine et al. (2013) use Fermi-LAT detected
GRB afterglows to constrain the MF decay in large scale. Pe'er
\& Zhang (2006) consider the MF decay effect on the GRB prompt emission,
and argue that for short enough MF decaying length scale, the
electrons become slow cooling, avoiding the strong fast cooling
problem in GRB spectrum. Derishev (2007) also points out that MF
decay behind the shock brings flexibility for the fast cooling
spectrum. Recently, Uhm \& Zhang (2013) also introduce the MF decay
to solve the fast cooling problem, but in a picture different from
the internal shock model.

Although Pe'er \& Zhang (2006) have pointed out that the extreme
fast cooling slope can be avoided with MF decay, one still needs to
further carefully consider how the synchrotron spectrum changes with
a varying MF structure, as well as the role of IC cooling.
In this paper, we carry out numerical calculation to study the effect of the
decay of downstream MF on the electron cooling process and on
shaping the GRB spectrum. In \S2, the model of electron cooling in
decaying MF is presented; \S3 is the analytical analysis of the
synchrotron spectrum; the results of numerical calculation of the
time-integrated synchrotron and IC spectra are shown in \S4; and \S5
is conclusion and discussion.

\section{Model}
We consider the internal shock model for GRBs. When two parts of the
GRB ejecta with different velocities collide, shock waves are
generated and propagate into the unshocked ejecta. Electrons are
accelerated at and near the shock front, and then produce
synchrotron and IC radiation during flowing downstream. We assume
the shock produced MF decay with distance away from the shock front.
The exact MF structure downstream is unclear, but as some authors
did (Lemoine, 2013; Medvedev \& Spitkovsky 2009), we take the
following two possible MF structure downstream. One is a
power law decay (PLD) with time, where the magnetic field in the
rest frame of the downstream plasma is
\begin{equation}
   B=
 \left\{\begin{array}{ll}B_0 &t\leq t_B\\
   B_{0}(t/t_B)^{-\alpha_B} &t>t_B
 \end{array} \right.~~~~~~~{\rm (PLD)},
\end{equation}
and the other is an exponential decay (ED) with time,
\begin{equation}
   B= B_{0}\exp(-t/t_B)~~~~~~~~~~{\rm (ED)}.
\end{equation}
Here $t$ is the time measured in the rest frame of the downstream
plasma since the entry at the shock front. The values of the
constants $B_0$ and $t_B$ are presented below.

Consider a GRB with observed luminosity $L$ and variability time
$\delta t$, and assume that the bulk Lorentz factor is $\Gamma$ and
that the fraction of internal energy carried by accelerated
electrons is $\epsilon_e$. The internal shock radius is estimated to
be $r=2\Gamma^2c\delta t$, and the electron number density (or proton number density) in the
rest frame of the outflow is given by $n_e=L/\Gamma^24\pi r^2
m_pc^3\epsilon_e$.

The MF generated by the shock is assumed to carry a
fraction $\epsilon_B$ of the postshock internal energy, thus the
postshock MF at the shock front is
$B_0=\sqrt{8\pi\epsilon_Bm_pc^2n_e}=5\times10^{4}L^{1/2}_{52}\delta
t_{-2}^{-1}
(\frac{\Gamma}{300})^{-3}(\frac{\epsilon_e}{0.3})^{-1/2}(\frac{\epsilon_B}{0.3})^{1/2}
$G, where the convention $Q=10^xQ_x$ is used.

If an electron with injection Lorentz factor $\gamma_m$ only cools by
synchrotron radiation in the magnetic field of $B_0$, the
synchrotron cooling time is $\tilde{t}_{c}=6\pi
m_ec/\sigma_T\gamma_mB_0^2=3\times10^{-4}L_{52}^{-1}\delta
t_{-2}^2(\f{\Gamma}{300})^6(\f{\epsilon_e}{0.3})(\f{\epsilon_{B}}{0.3})^{-1}
\gamma_{m,3}^{-1}$s. This is much shorter than the outflow dynamical
time, $t_{dyn}\simeq r/c\Gamma=10(\Gamma/300)\delta t_{-2}$s, but
much longer than the downstream plasma time scale,
$\omega_p^{-1}=(4\pi n_e q_e^2/m_p)^{-1/2}=1.6\times10^{-9}
L_{52}^{-1/2}\delta
t_{-2}(\frac{\Gamma}{300})^{3}(\frac{\epsilon_e}{0.3})^{1/2}$s. We parameterize the MF
decay time $t_B$ by
 \bea
\tau_B\equiv t_B/\tilde{t}_{c}.
 \eea
The Fermi shock accelerated electrons are expected to be a power law
distribution, $dn_e/d\gamma_e\propto \gamma_e^{-p}$, where
$\gamma>\gamma_m$ and $p\approx2.3$. To reconcile with the observed
peak energy in the sub-MeV band, the minimum electron Lorentz factor
should be $\gamma_m\sim10^3$. In the numerical calculation
we will take
\begin{equation}
  \gamma_m=10^3 ~{\rm and}~B_0=5\times10^4{\rm G}.
\end{equation}
The simulation of Keshet et al. (2009) indicated
$\tau_B>10^4\omega_{p}^{-1}/\tilde{t}_{c}\sim 0.1$. In our numerical calculation
here, we will take the value of $0.1\le\tau_B<5$. The MF decaying
slope in the PLD case, $\alpha_B$, is unclear, and here we adopt nominal
values of $0.5<\alpha_B<3$, including the values implied by
numerical simulations (Chang et al. 2008).

Since we are interested in the time-integrated emission during the
electron cooling, we consider impulsive injection of high energy
electrons at the shock front, and the electrons undergo synchrotron
and IC cooling when being carried away downstream from the shock
front. The evolution of electron energy distribution can be solved
using the continuity equation,
 \bea
 \f{\partial (dn_{e}/d\gamma_e)}{\partial t}+\f{\partial
   [\dot{\gamma_e}(dn_{e}/d\gamma_e)]}{\partial\gamma_e}=0.
 \eea
The initial energy distribution of electrons follows a power law
with particle index of $-p$. The time $t$ is measured in the rest
frame of the downstream plasma, starting from the injection at the
shock front. When the electrons are advecting downstream they
encounter a decaying MF, where the initial MF strength at injection is
$B_0$. Note that we neglect the adiabatic energy loss in the
continuity equation, which is valid given that the radiative cooling
is much faster than the expansion, as seen by $\tilde{t}_{c}\ll t_{dyn}$.

The radiative energy loss of the electrons can be described by
\begin{equation}
  \dot{\gamma_e}m_ec^2=-(P_{syn}+P_{IC})=-\f{4}{3}\sigma_T c\gamma_e^2
  \f{B^2}{8\pi}(1+Y),
\end{equation}
where $Y=P_{IC}/P_{syn}$ is the Compton parameter, and should be
function of time. For simplicity, we assume that the synchrotron photon
energy density that the electrons encounter during the cooling process
is independent of time $t$. This assumption is valid based on the
following arguments. Although we consider, from technical point of
view, impulsive injection of electrons in the calculation, the
injection in reality happens in a finite duration, in which the
shock wave crosses the colliding ejecta shell. The electrons encounter
photons emitted both by earlier and later injected electrons.
Moreover, the photon energy density at a certain position is
contributed by photons emitted from the whole emission region. Thus,
the synchrotron photon energy density is more or less constant,
i.e., independent of the distance from the shock front. We will also
neglect the Klein-Nishina (KN) effect in the IC scattering, and only
use Thompson scattering cross section in deriving the electron
cooling rate $\dot{\gamma}$. The KN effect gives a marginal correction
for injection electrons around $\gamma_m$, and it is even less
important when electrons cool down to lower Lorentz factors. The KN
effect may have stronger effect on electrons injected with much
higher energy, affecting the high energy photon index $\beta$, which
is not the focus in this work. Therefore, we take
\begin{equation}
  Y=\frac{u_{syn}}{u_B}=Y_0\left(\frac{B_0}{B}\right)^2,
\end{equation}
where $Y_0=u_{syn}/(B_0^2/8\pi)$, denoting the initial ratio of IC
to synchrotron power at shock front. We use the values of
$Y_0=0.5-5$ in the numerical calculation.

From the above model we can solve the electron energy distribution
evolving with time, and hence calculate the time-integrated
synchrotron spectrum. In order to calculate the IC spectrum, one
needs the energy distribution of the seed photons. Here the energy
density of the seed photons, for which we only consider the
synchrotron photons, is given by $Y_0$ and $B_0$, i.e.,
$u_{syn}=Y_0B_0^2/8\pi$. The spectral shape of the seed photons,
based on the arguments and assumption above, is approximated by
the time-integrated synchrotron photon spectrum.

\section{Analytical considerations}

Before carrying out the numerical calculation, let us analyze
analytically the low energy spectral index in the extreme cases.

Consider first the usual homogeneous MF case. We can approximate the
simultaneous synchrotron spectrum from a single electron as a
$\delta$ function at the characteristic synchrotron frequency
$\nu(\gamma_e)\propto\gamma_e^2$ at any given time during the
electron cooling. Since the cooling time is
$t_{c}\propto\gamma_e^{-1}$, the time-integrated energy spectrum
should be $\nu F_\nu t_{c}\propto \gamma_e\propto \nu^{1/2}$, which
spans between the cooling frequency and the injection frequency.

Next consider the MF decay cases. Since the ED case is not trivial
to analyze analytically, we consider here the PLD case. The electron
is subject to both synchrotron and IC cooling. For simplicity we
further separate the PLD case into two approximate regimes, the
synchrotron-only and the IC-only cases.

For the synchrotron-only case, we have
$\frac{d\gamma_e(t)}{dt}\propto{\gamma_e}(t)^2B(t)^2\propto
{\gamma_e}(t)^2t^{-2\alpha_B}$, and hence
\begin{equation}
      \gamma_e\propto \left\{\begin{array}{ll}
    t^{2\alpha_B-1}  ~~&\alpha_B<1/2\\
    {\rm const.}   ~~&\alpha_B>1/2
  \end{array} \right..
\end{equation}
 For $\alpha_B>1/2$, the MF decays too fast for the electron to
cool, which is less interesting for GRB prompt emission since huge
energy budget would be required. The characteristic synchrotron
peak frequency $\nu\propto \gamma_e^2 B(t)$ and the time ($\Delta
t$) within which the electrons mainly emit synchrotron photons at
$\nu$ are
\begin{eqnarray}
  \nu\propto \left\{\begin{array}{ll}
    t^{3\alpha_B-2}  ~~&\alpha_B<1/2\\
    t^{-\alpha_B}   ~~&\alpha_B>1/2
  \end{array} \right. ,\\
  \Delta t\propto \left\{\begin{array}{ll}
    \nu^{\frac{1}{3\alpha_B-2}} ~~ &\alpha_B<1/2\\
    \nu^{-\frac{1}{\alpha_B}}  ~~&\alpha_B>1/2
  \end{array} \right..
\end{eqnarray}
The time-integrated energy flux is then
    \begin{eqnarray}\label{eq:alpha1}
      \nu F_\nu \Delta t\propto
      \left\{\begin{array}{ll}
        \nu^{\frac{2\alpha_B-1}{3\alpha_B-2}} ~~& \alpha_B<1/2 \\
        \nu^{2-\frac{1}{\alpha_B}}~~ & \alpha_B>1/2
      \end{array}
    \right.,
    \end{eqnarray}
where $F_\nu\propto B$ is used. Note the spectrum is harder
than a slope of $4/3$ if $\alpha_B>3/2$, which is impossible for
synchrotron spectrum. This is due to the $\delta$ function
approximation of the synchrotron spectrum. Thus, in this case we
should use $\nu F_\nu\Delta t\propto\nu^{4/3}$.

For the IC-only case,
    \begin{equation}
    \gamma_e\propto t^{-1},~~
    \nu\propto \gamma_e^2 B\propto t^{-(2+\alpha_B)}, ~~
  \Delta t\propto \nu^{-\frac{1}{2+\alpha_B}}.
    \end{equation}
The time-integrated synchrotron spectrum is
    \bea\label{eq:alpha}
  \nu F_\nu \Delta t &\propto& \nu^{\frac{2\alpha_B+1}{\alpha_B+2}}.
  \eea
Note the spectrum is harder than a slope of $4/3$ if
$\alpha_B>5/2$, in which we should also use $\nu F_\nu\Delta
t\propto\nu^{4/3}$.

From equations (\ref{eq:alpha1}) and (\ref{eq:alpha}) (also see Lemoine 2013 and Derishev 2007 for similar derivations), one finds
that if $\alpha_B=0$, we recover the usual spectrum fast
cooling spectral slope of $1/2$. However, if the MF decays,
$0<\alpha_B<2/3$, the spectrum in the synchrotron only case becomes
softer than 1/2 (eq [\ref{eq:alpha1}]); while, on the contrary, in
the IC-only case the spectrum becomes harder than 1/2, and the larger
$\alpha_B$ the harder the spectrum (eq [\ref{eq:alpha}]). Thus, IC
cooling is important in producing a hard spectrum.

\section{Numerical results}

We present here our numerical results for the GRB spectra. We
take the case with the parameter values of $\alpha_{B}=1.5$,
$\tau_B=1.0$, $Y_0=0.5$ and $p=2.3$ as the fiducial model, which
gives a time-integrated synchrotron spectrum consistent
with the typical GRB spectrum with $\alpha\sim-1$ ($\nu
F_\nu\propto\nu$). The calculation is carried out in the rest frame
of the plasma downstream, but the resultant spectra have been
plotted in the observer frame using a typical bulk Lorentz factor
$\Gamma=300$.

\begin{figure}[t]
\includegraphics[width=\columnwidth]{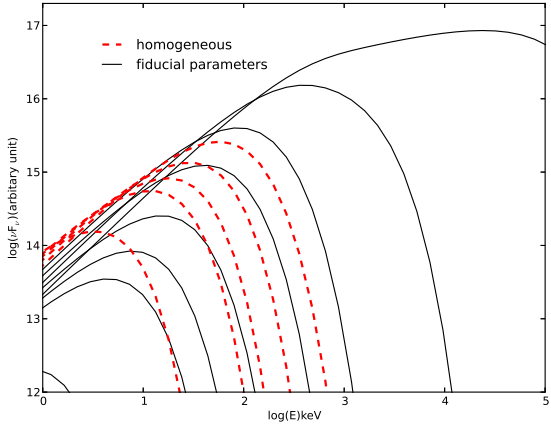}
\caption{The instantaneous synchrotron spectra for the PLD and
homogeneous MF cases. The fiducial parameters taken to be
$\Gamma=300$, $p=2.3$, $Y_0=0.5$, $\alpha_{B}=1.5$ and $\tau_B=1.0$.
The lines from right to left correspond to time $t=1\times
10^{-5}$s, $1\times 10^{-4}$s, $3\times 10^{-4}$s,
$4\times 10^{-4}$s, $6\times 10^{-4}$s, $8\times
10^{-4}$s, $1\times 10^{-3}$s, and $2\times 10^{-3}$s,
respectively. Note that the dashed and the solid lines are
superposed together for the times
$<4\times10^{-4}$s.}\label{fig:t-reso}
\end{figure}

\begin{figure}[t]
\includegraphics[width=\columnwidth]{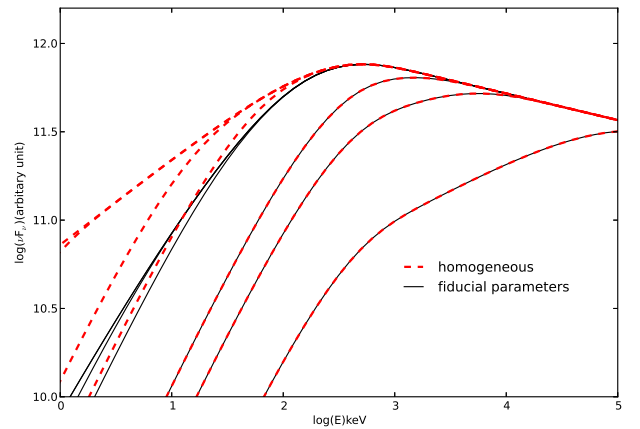}
\caption{The time-integrated synchrotron spectra up to
different times. The lines from right to left correspond to time
$t=1\times 10^{-5}$s, $5\times 10^{-5}$s, $1\times 10^{-4}$s,
$5\times 10^{-4}$s, $1\times 10^{-3}$s, $5\times 10^{-3}$s, and
$1\times 10^{-2}$s, respectively. Note that for $t>5\times 10^{-3}$s, the spectra are unchanged with time and superposed together in both the PLD and homogeneous MF cases. And the dashed and solid lines are superposed together for $t<5\times 10^{-4}$s.}\label{fig:t-timeinte}
\end{figure}

First, we show in Fig \ref{fig:t-reso} the instantaneous synchrotron
spectra of the injected electrons. One can find that the MF decay
case generally produces much weaker and harder synchrotron emission
than that in the homogeneous MF case at the same time $t$. Thus the
time-integrated spectrum should be harder. Then we calculate the
time-integrated spectra up to different times (Fig
\ref{fig:t-timeinte}) to illustrate how a harder spectrum in the low
energy band is formed.  From Fig \ref{fig:t-timeinte}, we can see
that in the MF decay case the low energy index becomes softer with
time and stabilizes at around $\sim1$ at late time, harder than that
in the homogeneous MF case.

\begin{figure*}[t]
\includegraphics[width=\textwidth]{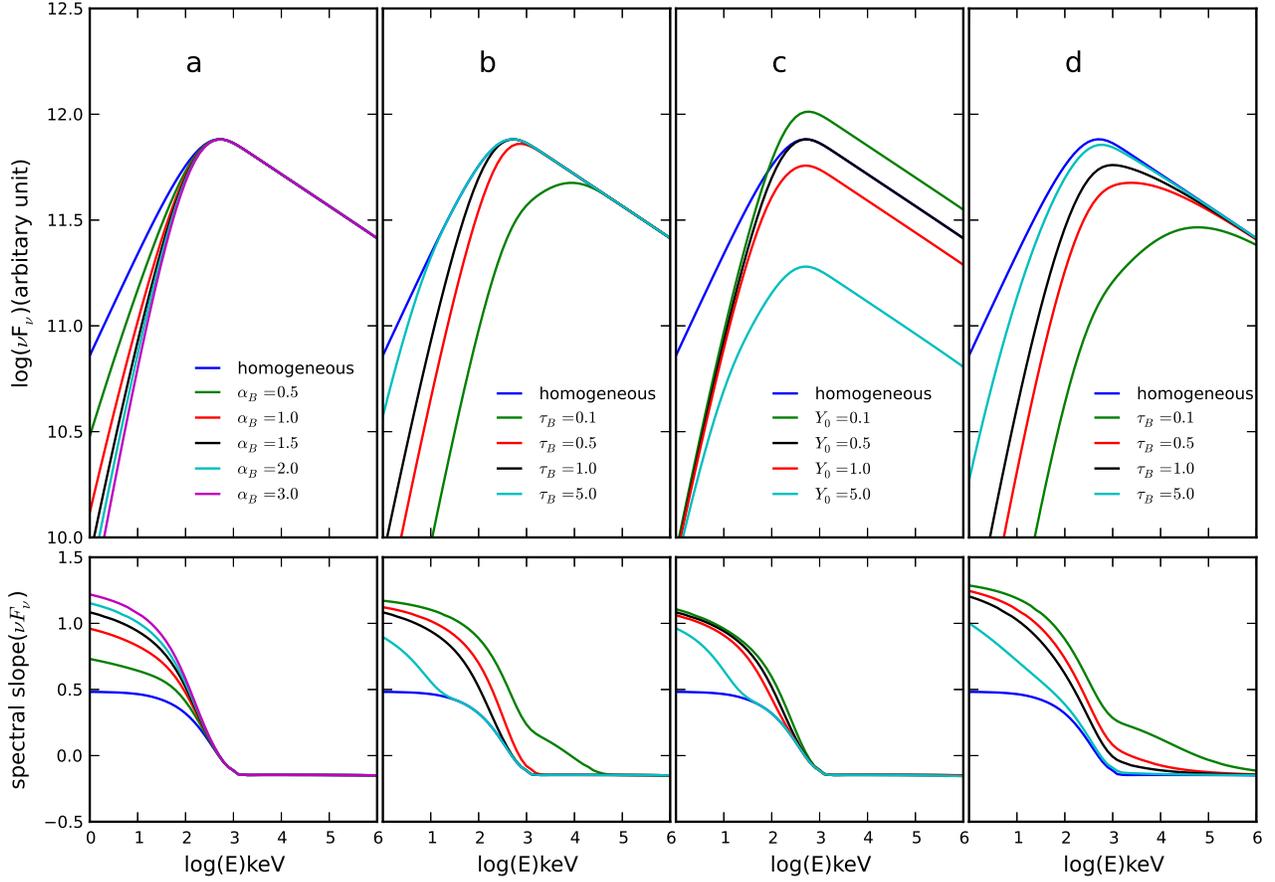}
\caption{Time-integrated synchrotron spectrum. The label
"homogeneous" indicates the homogeneous MF case. Panels a, b and c
correspond to the PLD case, while panel d is the ED case. The bottom
panels show the spectral slope as function of photon
energy.}\label{fig:timeinteg}
\end{figure*}

In Fig \ref{fig:timeinteg}, we show the time-integrated synchrotron
spectrum, calculated over a duration from the electron injection up
to $0.05$~s, which is about $\sim10^2\tilde{t}_{c}$. Up to this
time, the time-integrated spectrum does not vary any longer in the
interesting energy range and becomes a "steady" state. One can see
that the MF decay leads to harder low energy spectral slopes,
compared with the homogeneous MF case. By changing the parameter
values we can see how the resulted synchrotron spectrum varies.

From Fig \ref{fig:timeinteg} it can be seen that for the PLD case the low energy (below
injection frequency) spectrum is mostly sensitive to the MF decay
slope, $\alpha_B$; the spectrum is harder for larger $\alpha_B$. For
$\alpha_B$ approaching zero, the $\nu F_\nu$ spectral slope is close
to the homogeneous MF case, $1/2$; if $\alpha_B\ga2$ the slope is
close to the slow cooling slope, $4/3$. This is consistent with
predicted in equation (\ref{eq:alpha}).

Fig \ref{fig:timeinteg} also shows that in the PLD the MF decay time scale and the Compton
parameter do not sensitively affect the low energy slope. If the
decaying time is larger, i.e., larger $\tau_B$, the spectrum is
close to the homogeneous MF case, but the spectral slope in the
lowest energy range does not change much. Similarly, it can be seen that
changing the Compton parameter $Y_0$ does not change much the
spectral slope at the lowest energy end, while changing the normalization
of the synchrotron spectrum.

In the ED case, the spectrum also becomes harder than the
traditional homogeneous MF case, but similarly to the PLD case, the spectral
slope tends towards $4/3$ and does not change much with varying MF
decay time scale.

\begin{figure}[t]
\includegraphics[width=\columnwidth]{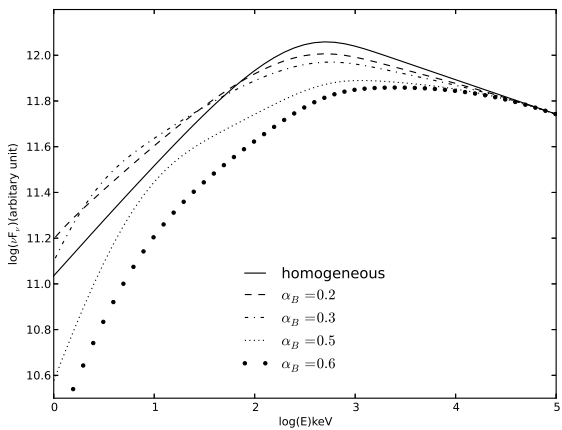}
\caption{The PLD case with synchrotron spectra softer than $\nu
F_\nu\propto \nu^{1/2}$. The $\alpha_B$ values are marked in the
plot, and $\tau_B=0.1$ and $Y_0=0$ are used. The other parameters
not mentioned are taken to be the fiducial values.
}\label{fig:integ-soft}
\end{figure}

We also calculate the case of a spectrum softer than 1/2, with
$0<\alpha_B<2/3$ and without IC cooling (Fig. \ref{fig:integ-soft}).
These represent a small fraction of burst cases (Preece et al.
2000). From Fig \ref{fig:integ-soft}, we can see that our numerical
calculations indeed obtain a softer spectrum, roughly consistent
with the prediction in equation (\ref{eq:alpha1}).

\begin{figure}[t]
\includegraphics[width=\columnwidth]{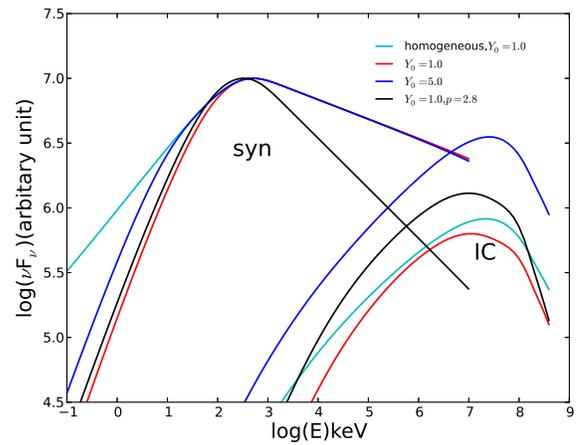}
\caption{Time-integrated synchrotron and IC spectra for the PLD case. The parameters are marked on the plot, other parameters not mentioned having their fiducial values. }\label{fig:IC}
\end{figure}

Finally, we calculate the time-integrated IC spectrum, shown in Fig. \ref{fig:IC}.
The IC spectral slope in the MF decay case in low energy band depends on
the low energy part of the synchrotron spectrum, so it is also harder than
the homogenous MF case. However, as in the case of a homogeneous MF, the
IC component is usually dominated by the synchrotron component even in the high
energy range, if the Compton parameter is not much larger than unity. This
is mainly due to the KN suppression of the IC emission. The IC
component in the MF decay case is even lower than that in the homogenous
MF case, because of the fact that in the former case there are more
soft seed photons that suffer less KN suppression.

The IC component only shows up if the Compton parameter is much
larger than unity (see the case of $Y_0=5$), or the injected
electron energy distribution is soft (so that the high energy
synchrotron spectral tail is soft; see the case of $p=2.8$). In
Fermi-LAT observations, an IC component is not explicitly confirmed in
most GRBs. But there are several GRBs showing an extra high energy
component (Abdo et al. 2009a, 2009b), which might be the IC contribution.

\section{Discussion and conclusion}

In this paper, we have assumed that the internal shock generated MF
decays with distance from the shock front, and taking into account the
electron cooling we calculated the synchrotron and IC emission. We
find that the synchrotron spectrum at low energies (below the injection
frequency) can be harder in this scenario than in the traditional scenario with
a homogeneous postshock MF. The observed GRB spectra with typical low energy
slopes $\alpha\sim-1$ are best reproduced in our power-law decay (PLD)
models when (i) the MF decay time is comparable to electron cooling time
($\tau_B\sim1$), (ii) the IC cooling is not much weaker than
synchrotron cooling ($Y_0\sim0.5$), and (iii) the temporal power-law
decay index is $\alpha_B\sim1.5$; or in our exponential decay (ED) models with
$\tau_B\sim1$. The low energy spectral index is most sensitive to
the MF decay index in the PLD case. We also find that the spectrum would be softer than
$\nu F_\nu\sim1/2$ if the MF decays slope $0<\alpha_B<2/3$ and the
IC cooling is negligible. These cases correspond to the marginal fast
cooling cases (or even slow cooling for $1/2<\alpha_B<2/3$) and thus are
less interesting. However, in the observations, a handful of bursts indeed
have spectra softer than 1/2 (Preece et al. 2000). So these MF decay
cases producing softer spectra might correspond to these observed bursts.
Our results suggest that the low energy slopes in the MF decay cases are
not fixed values of $\alpha=-3/2$ or $\alpha=-2/3$ (corresponding to fast
or slow cooling, respectively) any more, but vary from $-2<\alpha<-2/3$
($\nu F_\nu\propto\nu^{0-4/3}$), which accommodates the observations for
the vast majority of GRBs.

In the present MF decay cases, the synchrotron characteristic frequency
moves more rapidly towards lower frequencies, compared with the
homogeneous MF case. The energy which an electron emits at a certain
synchrotron frequency is much less due to the decrease of the
synchrotron cooling rate. The IC cooling becomes relatively stronger
as the synchrotron emission weakens. Subsequently, this
generates a harder spectrum than that in the homogeneous MF case.
However, the spectrum is unlikely to be harder than $\nu
F_\nu\propto\nu^{4/3}$. Thus, for the relatively small fraction
of observed GRBs with $\nu F_\nu$ low energy spectral index $>4/3$,
different effects may be at play, e.g. strong synchrotron self absorption,
or photospheric emission (e.g., Rees \& M\'{e}sz\'{a}ros 2005; Pe'er et al. 2006;
Beloborodov 2010).

The IC radiation will generate an extra component in the high energy band.
However, it is dominant only in the cases when the initial Compton parameter
is large, $Y_0\gg1$. Since the observations do not show a significant
high energy component for most GRBs, a high $Y_0$ is not the general
case. A very soft electron distribution is conducive to the emergence
of the IC component. Several Fermi GRBs showing an high energy
component do have a soft spectral slope $\beta\la-2.5$ above $E_p$
(Abdo et al. 2009a, 2009b), suggesting a soft electron energy
distribution.

Pe'er \& Zhang (2006) consider the MF decay effect on GRB
prompt emission in the internal shock model, assuming a sharp MF
decay. To avoid the fast cooling problem, the cooling frequency is
required to be around the typical synchrotron frequency
($\nu_c\sim\nu_m$). Thus the spectrum in the low energy bands in
their model would be $\nu F_\nu\sim\nu^{4/3}$. In the present paper,
we find that the MF decay with different power laws can generate a wide
range of spectra with $\nu F_\nu\propto\nu^{0-4/3}$, which
accommodates the data of the vast majority of GRBs. We also find that the
IC cooling, although unimportant in the total energy loss of
electrons, is important in shaping the spectrum in the low energy
bands.

Uhm \& Zhang (2013) recently have also considered the effect of an
MF decay, and produce hard synchrotron spectra consistent with GRB
observations. However, there are essential differences between their
model and ours. Their MF decay time is comparable to the dynamical
time of the outflow, $B\propto r^{-b}$ with $b\approx1$, and their
calculation keeps injecting electrons over a time which is longer by
many orders of magnitude than the dynamical time at the point when
the injection starts. Thus, the bulk electrons are injected at
larger radii $r\sim10^{16}$cm (and hence small $B$ and large cooling
time) so they do not cool in the extremely fast cooling regime. In
our work, by contrast, we have considered the internal shock model,
and the MF decay time is smaller  by many orders of magnitude than
the dynamical time. The electron cooling time is required to be
comparable to the MF decay time in order to avoid extreme fast
cooling, and the synchrotron spectrum is further shaped by the
suppression of synchrotron due to increasing IC cooling.

To reproduce the prompt GRB spectra in our model, the MF decay time should be comparable
to the electron cooling time, $\tau_B\sim1$, indicating a MF decay length much larger
than the plasma skin depth, typically $ct_B\sim10^5c/\omega_p$ (see also Pe'er \& Zhang
2006). This is different from the afterglow shock case (Gruzinov \& Waxman 1999;
Gruzinov 2001; Lemoine et al. 2013). For the prompt emission, if the GRB outflow
is magnetized before the shock, this might lead to a relevant scale which is not the
plasma scale but a much larger one.

\begin{acknowledgements}
We thank P. Veres, K. Kashiyama, D. Burrows, D. Fox, S. Gao, L. Uhm
and B. Zhang for helpful discussions and the anonymous referee for 
helpful comments. We acknowledge partial support
by the Chinese Natural Science Foundation (No. 11203067), Yunnan
Natural Science Foundation (2011FB115) and the West Light Foundation
of the CAS (XHZ); the NSFC (11273005), the MOE Ph.D.  Programs
Foundation, China (20120001110064), the CAS Open Research Program of
Key Laboratory for the Structure and Evolution of Celestial Objects
and the National Basic Research Program (973 Program) of China (Grant No. 2014CB845800)"
(ZL); NASA NNX13AH50G (PM, XHZ);
NSFC (No. 11133006) (JMB, XHZ); and NSFC (No. 11003014/A0303) (XWL)
\end{acknowledgements}

\bibliographystyle{plain}

\end{document}